\documentstyle[twoside,fleqn,espcrc2]{article}

\newcommand{\bea}{\begin{eqnarray}}
\newcommand{\eea}{\end{eqnarray}}
\newcommand{\be}{\begin{equation}}
\newcommand{\ee}{\end{equation}}
\newcommand{\bt}{\begin{tabular}}
\newcommand{\et}{\end{tabular}}
\newcommand{\ba}{\begin{array}}
\newcommand{\ea}{\end{array}}

\newcommand{\vev}[1]{\langle #1 \rangle}
\newcommand{\tr}{{\rm Tr}}
\newcommand{\pf}{{\rm Pf}}

\newcommand{\drawsquare}[2]{\hbox{%
\rule{#2pt}{#1pt}\hskip-#2pt
\rule{#1pt}{#2pt}\hskip-#1pt
\rule[#1pt]{#1pt}{#2pt}}\rule[#1pt]{#2pt}{#2pt}\hskip-#2pt
\rule{#2pt}{#1pt}}
\newcommand{\fun}{\raisebox{-.5pt}{\drawsquare{6.5}{0.4}}}
\newcommand{\funb}{\overline{\fun}}
\newcommand{\as}{\raisebox{-3.5pt}{\drawsquare{6.5}{0.4}}\hskip-6.9pt%
        \raisebox{3pt}{\drawsquare{6.5}{0.4}}}
\newcommand{\asb}{\overline{\as}}

\def \nonpert{non\discretionary{-}{}{-}per\-tur\-bative\ }

\def\Gpert{G_{\rm pert}}
\def\WNP{W^{\rm (NP)}}
\def\fNP{f^{\rm (NP)}}
\def\GNP{G_{\rm NP}}
\def\Mpl{M_{\rm Pl}}

\newcommand{\AmS}{{\protect\the\textfont2
  A\kern-.1667em\lower.5ex\hbox{M}\kern-.125emS}}

\title{Singularities in $d=4$,  $N=1$ Heterotic String Vacua
\thanks{Talk given by J.\ Louis at the conference
``Duality Symmetries in String Theory - II''
at the ICTP, Trieste, April 1-4, 1997 and at 
``Strings '97'' in
Amsterdam, June 16-20, 1997.}}

\author{Matthias Klein\address{Universit\"at M\"unchen, 
        Sektion Physik,\\ Theresienstr.~37, D-80333 M\"unchen,
        Germany}$^{\rm\scriptstyle ,b}$ 
        and Jan Louis\address{Martin--Luther--Universit\"at 
        Halle--Wittenberg,\\ FB Physik, D-06099 Halle, Germany}}

\begin{document}

\begin{abstract}
Singularities in the Yukawa and gauge couplings 
of $N=1$ compactifications of 
the $SO(32)$ heterotic string are discussed.
\end{abstract}

\maketitle

Perturbative $d=4$, $N=1$ heterotic vacua
are characterized by a $c=9$ conformal field
theory (CFT) with $(0,2)$ worldsheet supersymmetry
and the choice of a $\bar c = 22$
vector bundle. The space-time spectrum features
the gravitational multiplet, non-Abelian vector
multiplets, charged chiral matter multiplets $Q^I, \hat Q^{\hat I}$
and gauge neutral chiral moduli multiplets $M^i$.
The couplings of these multiplets
are described by an effective Lagrangian 
which is 
constrained by $N=1$ supersymmetry to only depend 
on three arbitrary functions: 
the real K\"ahler potential $K$, 
the holomorphic superpotential $W$ and the 
holomorphic gauge kinetic function $f$.
Due to their holomorphicity the latter two
obey a non-renormalization theorem \cite{LF}.
The superpotential $W$ receives no perturbative
corrections and one only has
\be
W=W^{(0)} + \WNP\ ,
\ee
where $W^{(0)}$ denotes the tree level contribution
while $\WNP$ summarizes  the \nonpert 
corrections. 
$W^{(0)}$ contains mass terms and 
Yukawa couplings both of which are 
generically moduli dependent. 
$\hat Q^{\hat I}$ are  the massive charged multiplets
of the string vacuum while $Q^I$ denotes the massless multiplets.
The massive modes $\hat Q^{\hat I}$ are 
commonly integrated out of
the effective Lagrangian since their typical mass is of order
of the Planck scale $\Mpl$. However, since their masses can be 
moduli dependent they might become light in special regions
of the moduli space. Therefore we choose
to keep such massive multiplets in the effective theory. Thus a generic 
tree level superpotential is given by
\bea
W^{(0)} &=& 
m_{\hat I}(M^i)\, \hat Q^{\hat I} \hat Q^{\hat I} \nonumber \\ 
&+& 
Y_{IJK}(M^i)\, Q^I  Q^J  Q^K + \ldots \ ,
\eea
where all gauge quantum numbers of 
$Q^I$ and $\hat Q^{\hat I}$ are suppressed.

The real part of the gauge kinetic function  $f$ determines the 
inverse gauge coupling according to
$g^{-2} = {\rm Re} f$.
The holomorphic $f$ receives no perturbative
corrections beyond one-loop and one  has
\be
f=f^{(0)}+f^{(1)}  + \fNP\ ,
\ee
where $f^{(1)}$ is the one-loop correction.

In most (if not all) heterotic string vacua
$W$ and $f$ are singular functions on the 
moduli space.\footnote{For a recent discussion
see ref.~\cite{KS}.}
For example, in a heterotic vacuum
obtained as a Calabi--Yau compactification
on  a quintic hypersurface in ${\bf CP^4}$ 
with  defining polynomial 
$p=\sum_{\alpha=1}^{5} X_\alpha^5 
- 5\psi X_1X_2X_3X_4X_5$
one finds \cite{CDGP,BCOV}
\be
Y \sim M^{-1} , \quad f \sim \log M\ ,
\ee
where $M\equiv 1-\psi^5$.
It is of interest to understand the 
physical origin of such singularities.
In this talk we expand on a suggestion by
Kachru, Seiberg and Silverstein \cite{KSS}
that non-perturbative gauge dynamics is the 
cause of singularities in $N=1$ heterotic vacua.

For type II vacua compactified on  
Calabi--Yau manifolds (or equivalently on 
$(c= \bar c=9)$ $(2,2)$ CFTs)
Strominger \cite{strominger}
gave a consistent picture of the physical
mechanism responsible for the singularities.
For such vacua the space-time effective theory is 
$N=2$ supersymmetric and, as a consequence,
there is only a logarithmic singularity
in the gauge couplings $g^{-2}$ but no power
like behaviour in any of the other couplings.
Specifically, for the quintic threefold one finds
\be
g^{-2} \sim  \log M\bar M + \ldots\ .
\ee

Such a correction to the gauge couplings
is induced in quantum field theories at one-loop 
by charged matter multiplets of  mass $M$.
Strominger suggested that there are indeed
\nonpert charged 
states in the type II 
string spectrum which cause the singularity in
the gauge couplings.
In perturbative string theory all such states 
are integrated out and one only sees their
effects in the moduli dependence of 
the gauge couplings.
However, as $M\to 0$ some of 
the \nonpert states become massless 
and it is no longer legitimate to integrate 
them out. Thus the singularity signals  light states
and the break down of an  effective low energy
theory which does not properly include all 
the light degrees of freedom.\footnote{As 
a consistency check it was shown 
that the \nonpert states also render the coupling  
of $R^2$ logarithmically singular at $M\to 0$ where
the coefficient of the singularity counts the number
of massless states \cite{vafa}.} 

The relevant non-perturbative states arise in $N=2$
hypermultiplets and they 
carry $U(1)$ charge of Ramond-Ramond vector bosons.
As a consequence, they do not have the canonical
couplings to the dilaton \cite{various}
but their mass only depends on
the scalar fields of the $N=2$ 
vector multiplets in the string spectrum.
Therefore, the corrections to the gauge couplings 
even though being non-perturbative appear without
any dilaton dependence and thus `compete' with
tree level effects.

In string vacua with $N=1$ supersymmetry
the situation is
more complicated. The power-like singularity
of the tree level Yukawa couplings
cannot be explained by states becoming
massless. Instead it was suggested
\cite{KSS} that 
at least some of the singularities
are caused by strong
coupling dynamics of an asymptotically free
\nonpert gauge group.
Such \nonpert gauge groups are known to 
arise at special points
in the moduli space of heterotic string vacua
in six space-time dimensions 
and 8 supercharges \cite{schmal}.
Such vacua can be constructed either as an 
abstract $(0,4)$ CFT or as geometrical $K3$
compactification of the $d=10$ heterotic string.
Consistency requires that the number of
instantons of the gauge group 
($E_8\times E_8$ or $SO(32)$)
is non-vanishing and equal to 24.
Associated with these instantons is a 
(quaternionic) moduli space 
which parameterizes their size, location 
and embedding into the gauge group.
This moduli space has singularities at the points
where instantons shrink to zero size.
The physical origin of these singularities 
is either a set of gauge bosons becoming
massless or an entire string becoming tensionless
\cite{schmal,hanany,SW,bele,aspinwall,intriligator,AM}.
Both of these effects are invisible in string perturbation theory;
they are \nonpert
in that they occur in regions of the moduli space
where string perturbation theory breaks down.

The moduli space  of $k$ $SO(32)$ instantons
is isomorphic to the Higgs branch
of an $Sp(2k)$\footnote{By $Sp(2k)$\ we mean the rank $k$\ 
symplectic group whose fundamental representation has 
dimension $2k$.} gauge theory with 32 half-hypermultiplets
 in the fundamental 
($\fun$) $2k$-dimensional representation,
a traceless antisymmetric tensor ($\as$) in the
$k(2k-1)-1$-dimensional representation and 
a singlet \cite{schmal}. 
The singularities of this moduli space
precisely occur where some (or all) of the 
$Sp(2k)$ gauge bosons become massless.

The situation is more involved when an $SO(32)$
instanton shrinks on a singularity of the
un\-derlying CFT or K3 manifold.
Here we only discuss the simplest case of
an $A_1$ singularity.
One has to distinguish between instantons
`with vector structure' and  
`without vector structure' \cite{bele}.\footnote{
This terminology refers to properties of the $SO(32)$
connection at infinity. See refs.~\cite{bele,aspinwall}
for details.}  
For instantons without vector structure
the moduli space is conjectured to be
isomorphic to the Higgs branch of a
$U(2k)$ gauge theory with 16 hypermultiplets
in the fundamental ($\fun$) $2k$-dimensional representation and
two antisymmetric tensors ($\as$) in the
$k(2k-1)$-dimensional representation \cite{BS,GP,DM,bele}.

Small instantons with vector structure on an $A_1$ singularity show a
more complicated behaviour 
\cite{aspinwall,intriligator,AM,DM}.\footnote{
We thank P.~Aspinwall and K.~Intriligator
for a useful correspondence on this point.}
The case where less than four instantons
coalesce on the singularity  is 
not fully understood yet
while the moduli space of four (and more)
instantons on the singularity
has a Higgs branch
and a Coulomb branch. On the Coulomb branch the dimension of the 
moduli space has been reduced by 29 but an additional tensor 
multiplet is present.

In this talk we focus on the $SO(32)$ heterotic string 
and only consider  the \nonpert  effects
associated with \nonpert gauge bosons.\footnote{
Singularities in the
$E_8\times E_8$ heterotic string 
and chirality changing phase transitions
have recently been discussed
in ref.~\cite{K-S}.}
In this case the gauge group $G$ of the string vacuum
is a product of the perturbative
gauge group $\Gpert$ and the \nonpert gauge group
$\GNP$
\be
G=\Gpert\times\GNP\ ,
\ee
where $\GNP$ is a subgroup of either $Sp(2k)$ or 
$U(2k)$.\footnote{By giving an appropriate 
vacuum expectation value to one of the two
antisymmetric tensors of $U(2k)$ one arrives
at the $Sp(2k)$ gauge theory with the precise
spectrum given above.}

Upon toroidally compactifying the above theories one obtains
$N=2$ string vacua in $d=4$. String vacua with $N=1$ 
supersymmetry arise when one compactifies not on $K3\times T^2$
but on a Calabi--Yau threefold. However, there is a particular class
of threefolds -- $K3$ fibrations --
which is closely related to the six-dimensional
heterotic vacua. $K3$ fibrations are
3-dimensional Calabi--Yau manifolds where a $K3$ is  fibred
over a ${\bf P^1}$ base. If the base is large the adiabatic argument
applies \cite{VW} and the singularities
of the $K3$ fibres are inherited from the corresponding
six-dimensional vacuum \cite{KSS}. 

A specific class of $K3$ fibrations
are the quintic hypersurfaces 
defined in weighted projective space
${\bf WP^4}_{1,1,2k_1,2k_2,2k_3}$ \cite{CDFKM,KLM}.
For compactifications of the $SO(32)$ heterotic string on
$(0,2)$ deformations of
such Calabi--Yau spaces the \nonpert 
spectrum is computed in ref.~\cite{KSS}
for the case of a single small instanton
at a smooth point in the $K3$ fibre.\footnote{It is important 
to consider a $(0,2)$ deformation since on the $(2,2)$ locus the spin
connection is embedded in the gauge connection and a small
instanton necessarily has to shrink on a $K3$ singularity.}
It is found that the \nonpert gauge group in the four-dimensional
vacuum is given by 
$\GNP = Sp(2) \cong SU(2)$ and out of the 32 half-hypermultiplets
in $d=6$ 
only four doublets in chiral $N=1$ supermultiplets 
(which we denote by $q_i, i=1, \ldots, 4$)
survive in $d=4$. 
This resulting gauge theory is asymptotically free 
($b_{SU(2)}>0$) and thus becomes
strongly coupled below its characteristic scale $\Lambda_{SU(2)}$.
It has the additional property that 
\be
c:= T({\bf ad}) - \sum_r n_r T(r) = 0\ ,
\ee
where $T(r)$ is the index in the representation $r$,
$T({\bf ad})$ is the index in the adjoint representation and
$n_r$ counts the number of chiral multiplets in 
representation $r$. 
(In this notation the one-loop coefficient
of the $\beta$-function is given by
$b=3T({\bf ad}) - \sum_r n_r T(r)$.)
The coefficient $c$ also appears in the anomaly 
equation of the R-symmetry. An anomaly free
R-symmetry implies  
$\sum_r n_rT(r)\, R_r = -c$,\ where
$R_r$\ is  the $R$-charge of the superfield.
As a consequence,  whenever $c=0$ 
one can choose $R=0$ for all superfields,
and hence
no \nonpert superpotential $\WNP$ 
can be generated by the strong coupling 
dynamics \cite{ADS,confine}. This conclusion is
believed to hold irrespective of the precise
form of tree level superpotential. 

Below $\Lambda_{SU(2)}$ the theory confines and the surviving
degrees of freedom are the gauge singlets 
\be
M_{ij} := q_i \cdot q_j\ .
\ee
$M_{ij}$ is antisymmetric and obeys the constraint
$\pf M:= {1\over 8}\epsilon^{ijkl} M_{ij} M_{kl} =0$; thus 
there are five physical degrees of freedom
in the effective theory.
Quantum mechanically the constraint
is modified and reads \cite{Seib_D49}
\be
\pf M= 
\Lambda_{SU(2)}^4 \ .
\ee
So altogether the superpotential is given by 
\bea
W &= &Y_{IJK}(M)\, Q^I Q^J Q^K 
+ m_{\hat I} (M)\, \hat Q^{\hat I} \hat Q^{\hat I}\nonumber\\
&&+ \lambda (\pf M-\Lambda^4) \\
&&+ \hbox{non-renormalizable\ terms}  \nonumber\\
&&+ \hbox{stringy non-perturbative terms}\ ,\nonumber
\eea
where $\lambda$ is a Lagrange multiplier.
Note that even though no non-perturbative 
superpotential is generated by the strongly coupled
gauge theory there are `stringy'
non-perturbative corrections of order
${\cal O}(e^{-g_{\rm string}^{-2}})$
where $g_{\rm string}$ is the string coupling 
constant or equivalently the vacuum expectation
value of the dilaton multiplet.

The non-renormalizable interactions typically include
mass terms for the $M_{ij}$. 
One way to see this
is to note that a corresponding $N=2$ vacuum
includes a chiral multiplet $\phi$ in the adjoint representation
with couplings (in $N=1$ notation)
\be
W = \kappa^{ij}\, q_i\cdot\phi\cdot q_j + \mu\, \tr \phi^2\ .
\ee
$N=2$ enforces $\mu=0$ and $\kappa^{ij} = \sqrt 2 \delta^{ij}$
while for $N=1$ $\phi$ becomes heavy and $\kappa$ arbitrary.
Integrating out $\phi$ and diagonalizing 
the resulting mass matrix leads to
\be
W = \frac12 \sum_{i,j} m_{ij} M_{ij}^2 \ .
\ee
(We choose $\Mpl=1$ throughout this talk.)
The number of moduli 
(or equivalently the dimension of the moduli 
space)  is 
given by $5$ minus the number of non-vanishing
mass terms $m_{ij}$.

Ref.~\cite{KSS} considered the case of a one-dimen\-sional
moduli space, that is a superpotential 
with four mass terms 
\bea
W &= &Y_{IJK}(M)\, Q^I Q^J Q^K 
+ m_{\hat I} (M)\, \hat Q^{\hat I} \hat Q^{\hat I}\nonumber\\
&&+ \lambda (\pf M-\Lambda^4)+ m_{13} M_{13}^2+ m_{14} M_{14}^2\nonumber\\
&&+ m_{23} M_{23}^2+ m_{24} M_{24}^2 + \ldots \ .
\eea
Integrating out the massive modes via 
$\frac{\partial W}{\partial M} =0$
results in 
\bea
M_{23}= M_{24}=M_{13}=M_{14} = 0 \ , \nonumber\\
M_{12}M_{34} = \Lambda^4\ .
\eea
(Note that in string theory $\vev{Q}=0$
holds by construction.)

In string perturbation theory both 
the Yukawa couplings $Y_{IJK}(M)$ and the masses 
$m_{\hat I} (M)$  are given as power series expansion
in the moduli. However, the strong coupling effects which are responsible
for generating the \nonpert constraint
$\pf M= \Lambda_{SU(2)}^4$ remove the origin of the moduli space
and render the perturbative expansion of the Yukawa couplings
singular. For example
\be
Y_{IJK}(M) \sim M_{12}+ M_{34} + \ldots
\ee
produces a singularity 
\be\label{singularity}
Y_{IJK}(M) \sim \frac{\Lambda^4}{M_{34}} 
+ \ldots \quad {\rm as\ }M_{34}\to 0\ .
\ee
Furthermore, the $SU(2)$ scale 
$\Lambda_{SU(2)}$ 
depends on the gauge coupling $g_{SU(2)}$  
of the \nonpert $SU(2)$
gauge bosons via 
$\Lambda_{SU(2)} \sim {\rm exp}({-\frac{8\pi^2}{bg^2_{SU(2)}}})$.
It was shown in refs.~\cite{sagnotti,DMW} that the gauge couplings 
of the \nonpert gauge bosons do not depend on $g_{\rm string}$ 
and hence the Yukawa couplings in eq.~(\ref{singularity})
do not depend on $g_{\rm string}$ either. 
Thus, exactly as in the $N=2$ case,
the non-perturbative effect which generates the 
singularity in eq.~(\ref{singularity})
does not have the standard dilaton or $g_{\rm string}$
dependence but instead competes with tree level
couplings. (Of course, this is a requirement for a consistent
explanation of the singularities in the Yukawa couplings.)

The mass terms  $m_{\hat I}$  either become large
or small at the singular points in the moduli space. 
We already learned from the $N=2$ analysis 
that light states 
($m_{\hat I} (M) \sim M_{34}$)
 which are charged under 
$\Gpert$ produce a singularity
in the associated gauge coupling at one loop;
they contribute a correction 
\be\label{FTsing}
g^{-2}= \sum_{\hat I} \frac{b_{\hat I}}{16\pi^2}\,
 {\rm log} |m_{\hat I}|^2 
+\ldots\ .
\ee
The coefficient of the singularity is set 
by the multiplicity
of the light modes and their gauge quantum numbers.
Unfortunately, in string theory
this coefficient is so far only known for
$(2,2)$ vacua \cite{BCOV,KL} and hence a more detailed comparison with the 
$(0,2)$ models considered in ref.~\cite{KSS} cannot be presented.
Conversely, it is not known at present how to repeat the 
analysis of ref.~\cite{KSS} for $(2,2)$ vacua since the 
\nonpert physics in the corresponding 
$d=6$ vacua is not fully understood.

Nevertheless, it is interesting to display the 
coefficients of the singularity in $(2,2)$ vacua.
One finds for  all 
$(2,2)$ vacua of the $SO(32)$ heterotic string, where
$\Gpert = SO(26)\times U(1)$, the constraint 
\be
16\pi^2 \big(g^{-2}_{SO(26)} 
- \frac{1}{6} g^{-2}_{U(1)}\big) = -16 F_1\ ,
\ee
while for $(2,2)$ vacua of the $E_8\times E_8$
heterotic string ($\Gpert = E_8\times E_6$)
one has
\be
16\pi^2 \big(g^{-2}_{E_8} - g^{-2}_{E_6}\big) = -12 F_1\ .
\ee
$F_1$ is the topological index defined in 
ref.~\cite{BCOV} which for the quintic hypersurface
in ${\bf CP^4}$ is given by
\be\label{Fone}
F_1 = - \frac{1}{12}\, {\rm log} M \bar M + \ldots \ .
\ee
One can easily check that   
eqs.~(\ref{FTsing})--(\ref{Fone}) are not easy
to satisfy.
One finds  only two sensible solutions for
$SO(26)\times U(1)$:\footnote{This was worked out jointly
with V.~Kaplunovsky.} either one has one pair of $SO(26)$ 
singlets with $U(1)$ charges $\pm 2$ or four pairs of 
$SO(26)$ singlets with $U(1)$ charges $\pm 1$.
(Here $m=M$ was assumed, $m$ being the mass of the light states;
if $m=M^2$ ($m=M^4$) one can also have two (one) pairs with
$U(1)$ charges $\pm 1$.)
In particular no states carrying $SO(26)$ charge 
can become light at the singularity.
Furthermore, there is no sensible solution for 
$E_8\times E_6$ at all.
This might indicate that in
$(2,2)$ vacua a different mechanism 
is responsible for the singularities 
in the gauge couplings as well as the Yukawa
couplings.\footnote{We checked that this conclusion
also holds for the singularities in the $K3$ fibre
of the two-parameter models of 
ref.~\cite{CDFKM} and we suspect that it is 
valid in general.}

In most Calabi--Yau vacua the singularity of the 
Yukawa couplings has a more complicated structure 
than just a simple pole.
In particular one observes generically a singular locus
with more than one component where the different
components can intersect in various ways.
Such a behaviour is reproduced by $k$ small instantons
located at the same point in moduli space or in other words
by $\GNP=Sp(2k)$. For this case the analysis of ref.~\cite{KSS} 
can be repeated without major modifications.
One finds that out of the 32 half-hypermultiplets 
in the fundamental $\fun$ representation 
of the six-dimensional vacua again
four chiral $\fun$ multiplets
remain in $d=4$. In addition, the antisymmetric tensor $\as$
survives as a chiral multiplet.
As before, the resulting spectrum in $d=4$ is 
an asymptotically free
gauge theory with the additional property that $c_{Sp(2k)} = 0$
for all values of $k$. Thus, no non-perturbative 
superpotential can be generated by 
the strongly coupled $Sp(2k)$ gauge theory.
Fortunately, this gauge theory has been analysed in some detail 
in ref.~\cite{CSS} (see also \cite{IP}). 
The physical degrees of freedom 
below $\Lambda_{Sp(2k)}$ are found to be
\bea
M_{ij}^l &:= &q_i \cdot A^l \cdot q_j, \quad i,j=1,\ldots,4,\nonumber\\
&&\qquad\qquad\qquad l=0,\ldots,k-1\ ,\nonumber\\
T_r &:= &\frac{1}{4r}\,  \tr A^r,\quad r=2,\ldots,k\ ,
\eea
where $A$ denotes the antisymmetric tensor.
As for the $SU(2)$ gauge theory the $M_{ij}^l$
and $T_r$ are not all independent but related by constraint equations.
These constraints are rather involved and for simplicity we
focus on $Sp(4)$ henceforth. For this case one has \cite{CSS}
\bea
T_2\, \pf M^0 + {\textstyle\frac12} \pf M^1 &= &2\Lambda^6\ ,\nonumber\\
 \epsilon^{ijkl} M_{ij}^0 M_{kl}^1 &= &0\ ,
\eea
where classically the left hand side of the first equation vanishes. 
Hence, there are  $2\cdot 6 + 1 -2 =11$ physical degrees of freedom
in the effective theory.
Exactly as in the previous $SU(2)$ example
these states can be massive and 
the superpotential has the generic  form
\bea\label{spfour}
W \hspace{-1em}&&= \ Y_{IJK}(M,T)\, Q^I Q^J Q^K 
+ m_{\hat I} (M,T)\, \hat Q^{\hat I} \hat Q^{\hat I}\nonumber\\
&&+ \lambda\ (T_2\, \pf M^0 +{\textstyle\frac12} \pf M^1 -2\Lambda^6)
\nonumber\\
&&+\mu\, {\textstyle\frac14}\epsilon^{ijkl} M_{ij}^0 M_{kl}^1\\
&&+\frac12 \sum_{i,j}\left( m^0_{ij}(M^0_{ij})^2+(m^1_{ij}(M^1_{ij})^2\right)
+m T_2^2\nonumber\ \\
&&+ \ldots\ .  \nonumber
\eea

The dimension of the moduli space depends
on the number of non-vanishing mass terms in 
eq.~(\ref{spfour}).
Consider, for example, the case $m^0_{12}=m^0_{34}=m=0$ 
which results in a two-dimensional moduli
space. 
The equations of motion are solved by
\bea
\lambda=\mu=M^1_{ij} &= &0\quad \forall i,j\nonumber\\
M^0_{13}=M^0_{14}=M^0_{23}=M^0_{24} &= &0\nonumber\\
T_2\, M^0_{12}\, M^0_{34} &= &2\Lambda^6\ .
\eea
A Yukawa coupling of the form
\be
Y_{IJK} \sim T_2+ M^0_{12}+ M^0_{34} + \ldots
\ee
now produces a singularity 
\be
Y_{IJK}(M) \sim 
\frac{2\Lambda^6}{M^0_{12}M^0_{34}} + \ldots 
\ee
as $M^0_{12}\to0$ or $M^0_{34}\to0$.
This is an example for a Yukawa coupling
which depends on two intersecting singular lines.

Different combinations of mass terms can lead to 
different properties of the singular locus.
For example, consider the case
$m^0_{12}=m^0_{34}=m^1_{12}=m^1_{34}=m=0$
for which the equations of
motion lead to
\bea\label{sol2}
\lambda=\mu &= &0\nonumber\\
M^0_{13}=M^0_{14}=M^0_{23}=M^0_{24} &= &0\nonumber\\
M^1_{13}=M^1_{14}=M^1_{23}=M^1_{24} &= &0\\
M^0_{12}M^1_{34}+M^1_{12}M^0_{34} &= &0\nonumber\\
T_2M^0_{12}M^0_{34}+{\textstyle \frac12}
M^1_{12}M^1_{34} &= &2\Lambda^6\ .\nonumber
\eea
If we solve the last two equations 
for $T_2$ and 
$M^0_{34}$, say, we find for a Yukawa coupling
of the form
\be
Y(M) \sim
T_2+M^0_{12}+M^0_{34}+M^1_{12}+M^1_{34}+\ldots
\ee
the singularities
\be
Y(M) \sim {(M^1_{12})^2\over 2(M^0_{12})^2}
-{2M^1_{12}\Lambda^6\over (M^0_{12})^2M^1_{34}}
-{M^0_{12} M^1_{34}\over M^1_{12}}\ ,
\ee
as $M^0_{12},\ M^1_{12}$ or $M^1_{34}\to0$. There are three singular
components but at most two of them appear in the same term of the
Yukawa couplings, i.e., only two of them are intersecting. If the
dimension of the moduli space equals the number of 
intersecting singular components then there cannot be more
than two singular lines in the $Sp(4)$ model.

Singular Yukawa couplings with more than two 
intersecting components can arise in
an $Sp(2k)$ gauge theory once we have $k>2$.
More precisely, in order to generate $l$
intersecting singular components we need
to look at a quantum constraint of the form 
\be
T_2\, T_3 \ldots T_l\, \pf M^0  + \ldots =\Lambda^{2k+2}
\ee
to find the minimal $k$ required. This yields
\be
4k\ge l(l+1)+2
\ee
as a necessary condition.
A proof of this relation together
with an extended analysis of the singularities
in the Yukawa couplings
will be presented elsewhere
\cite{klein}.

Finally let us consider the case where 
$\GNP=SU(2k)$. This gauge group arises when $k$
instantons without vector structure
shrink on an $A_1$ singularity of the $K3$ fibre.
Again the computation of the $d=4$ spectrum
can be performed following the methods 
of ref.~\cite{KSS}. 
One finds two flavours of fundamentals, i.e.\
$2(\fun + \funb)$,  as well as one flavour
of antisymmetric tensors $(\as + \asb)$.
As before, this an asymptotically free gauge 
theory with $c_{SU(2k)}=0$ for any $k$.
Consequently this theory confines 
below  $\Lambda_{SU(2k)}$ but no
non-perturbative superpotential is generated
by the strong coupling dynamics.

For simplicity we focus on $\GNP=SU(4)$ where
$\as\cong\asb$ holds. 
Therefore the two antisymmetric tensors $A_r,\ r=1,2,$ transform as a doublet 
under an additional $SU(2)$ flavour symmetry. 
The low energy degrees of freedom are 
found to be \cite{confine}
\bea
&&M^0_{ij}:=q_i\bar q_j,\quad i,j=1,2,\nonumber\\
&&M^1_{ij}:=q_i A^{2}\bar q_j\ ,\nonumber\\
&&H_r:= q A_r q,\quad r=1,2, \\
&&\bar H_r:= \bar q A_r \bar q\ ,\nonumber\\
&&T_{rs}:=A_rA_s\ .\nonumber
\eea
These singlet fields satisfy the 
additional constraints
\bea
&&\!\!\!\!\!\!\!\!\det T\det M^0 
- \epsilon^{rs}\epsilon^{tu}T_{rt}H_s\bar H_u - \det M^1
=\Lambda^8, \nonumber\\
&&\!\!\!\!\!\!\!\! \epsilon^{ij}\epsilon^{kl}M^0_{ik}M^1_{jl}
+\epsilon^{rs}H_r\bar H_s = 0\ .
\eea
The structure of the singularities again depends
on the mass terms  for the low energy degrees of freedom.
For example, mass terms 
for the fields $M^1, H, \bar H, 
M^0_{12}, M^0_{21}$
and $T_{12}$ lead to the equations of motion
$M^1=H= \bar H =
M^0_{12} = M^0_{21} =T_{12}=0$ and 
result in a 
three-dimensional moduli space satisfying the 
constraint
\be
T_{11}T_{22}M^0_{11}M^0_{22}=\Lambda^8\ .
\ee
In this case the singular locus of the Yukawa couplings 
has three intersecting components.
A more detailed analysis of the $SU(2k)$
gauge theories will also be presented in ref.~\cite{klein}.

\vspace{1cm}

\noindent {\bf Acknowledgement}

We would like to thank the organisers 
for hosting this exciting conference.

We have greatly benefited from discussions
and email correspondence with  P.\ Aspinwall,
T.~Banks, S.~F\"orste, K.~Intriligator, 
S.~Kachru, V.~Kaplunovsky,
T.~Mohaupt, G.~Moore, R.~Schimmrigk, 
M.~Schmaltz, E.~Silverstein, 
J.~Sonnenschein and S.~Theisen.

The work of M.K. is supported by the DFG. 
The work of J.L. is supported in part by GIF -- the German--Israeli
Foundation for Scientific Research.

\end{document}